# Local Cooperativity Mechanism in the DNA Melting Transition

Vassili Ivanov[1], Dmitri Piontkovski[2], and Giovanni Zocchi[1]

[1] Department of Physics and Astronomy, University of California Los Angeles, Los Angeles, CA 90095-1547, USA

[2] Central Institute of Economics and Mathematics, Nakhimovsky prosp. 47, Moscow 117418, Russia

We propose a new statistical mechanics model for the melting transition of DNA. Base pairing and stacking are treated as separate degrees of freedom, and the interplay between pairing and stacking is described by a set of local rules which mimic the geometrical constraints in the real molecule. This microscopic mechanism intrinsically accounts for the cooperativity related to the free energy penalty of bubble nucleation. The model describes both the unpairing and unstacking parts of the spectroscopically determined experimental melting curves. Furthermore, the model explains the observed temperature dependence of the effective thermodynamic parameters used in models of the nearest neighbor (NN) type. We compute the partition function for the model through the transfer matrix formalism, which we also generalize to include non local chain entropy terms. This part introduces a new parametrization of the Yeramian-like transfer matrix approach to the Poland-Scheraga description of DNA melting. The model is exactly solvable in the homogeneous thermodynamic limit, and we calculate all observables without use of the grand partition function. As is well known, models of this class have a first order or continuous phase transition at the temperature of complete strand separation depending on the value of the exponent of the bubble entropy.

# I. INTRODUCTION

Conformational transitions are a major area of research in the physics of biological polymers, and in this area DNA melting represents one model problem, because of its relative simplicity compared for instance to protein folding. The stable conformation of double stranded (ds) DNA at room temperature is the double helix. The two strands are held together by hydrogen bonds between Watson-Crick complementary base pairs (bp): A-T, stabilized by 2 hydrogen bonds ($\Delta G_{37}^o \sim 1.5\ k_B T_{310}$), and G-C, with 3 hydrogen bonds ($\Delta G_{37}^o \sim 3\ k_B T_{310}$) [1]. The next conformationally important attractive interaction is between adjacent bases on the same strand. This interaction favors keeping successive bases of the same strand stacked like a deck of cards, and is refereed to as stacking. The main destabilizing interaction is the repulsive electrostatic force between the negatively charged phosphate groups on either strand; this interaction can be modulated through the ionic strength of the solvent. As the temperature is raised, two processes alter the double helical conformation: base pairs may break apart, giving rise to bubbles of single stranded (ss) DNA, and bases may unstack along the single strands [2]. At the critical temperature where the two strands completely separate, there can still be significant stacking (i.e. secondary structure) in the single strands; this melts away as the temperature is raised further, the single strands finally resembling random coils at sufficiently high temperature.

Simplified statistical mechanics models of the transition (i.e. models with a reduced number of degrees of freedom compared to the real molecule) have a multiplicity of purposes, from quickly predicting melting temperature for applications such as PCR primer design, to studying the nature of the phase transition (whether continuous or discontinuous, for instance) and how it depends on sequence disorder, applied fields, etc. Indeed, to extract the thermodynamic parameters related to DNA melting from the experimental measurements requires a statistical mechanics model specifying the states in configuration space to which the free energies to be measured refer. For practical purposes, the nearest neighbor (NN) thermodynamic model [3] gives adequate predictions. In this model, the free energies for all possible dimer combinations (i.e. double stranded (ds) sequences of length 2) are assigned, and the total free energy is the sum of the dimer free energies for the specific sequence. Since there are 10 different dimers, this model has 10 parameter sets. Pairing and stacking interactions are lumped together into effective free energies of the dimer. Most theoretical investigations of the nature of the melting transition have on the other hand been carried out using models of the Poland-Scheraga (PS) kind or the Peyrard-Bishop (PB) kind. In the PS type models, the partition function for the molecule is written as a sum over bubble states [4–7]. A considerable effort has been devoted to the non trivial question of the correct statistical weights for the bubbles and the order of the phase transition in the thermodynamic limit [8,9]; by contrast, here we focus on the question of which degrees of freedom are crucial for a statistical mechanics description of oligomer melting curves. The PB approach is based on an effective Hamiltonian for the molecule which contains potential energy terms for the relative motion of opposite bases on the two strands as well as adjacent bases on the same strand [10]. This is closest to our own approach described below, in that here a local mechanism arising from the interplay of two different degrees of freedom is responsible for the cooperative behavior of (or long range interactions along) the bubbles. Modern developments of Peyrard-Bishop like Hamiltonian models were applied to describe DNA unzipping under an external force [11–14].

In a previous paper [15] we underlined the notion that melting profiles of DNA oligomers show the contribution of two different processes: unpairing of the complementary bases on opposite strands, leading to local strand separation, and unstacking of adjacent bases on the same strand, leading to loss of the residual secondary structure of the single strands, i.e. to a random coil conformation of the ss. Accordingly, if a model is to describe the melting profiles in the whole experimental temperature range, including at temperatures beyond strand separation, but where residual stacking may still be present in the ss, then pairing and stacking must be considered as separate degrees of freedom in the model. In [15] we pursued this approach through a model combining a description of stacking in terms of an Ising model and pairing in terms of a zipper model. We showed by comparison with the experiments that the Ising model gives an adequate description of stacking. However, the zipper model description of pairing [15], initially adopted for simplicity, is unsatisfactory in that it considers only states where the molecule unzips from the ends. Here we introduce an improved description, where pairing and stacking are both Ising variables, there are only nearest neighbor interactions, but there are constraints on the possible states of the fundamental dimer (two adjacent bp), which represent the geometrical constraints in the real molecule. We detail the states of the dimer below, but to give an example, one open bp followed by one closed bp means necessarily at least one unstacking, translating into one out of all possible dimer states being forbidden. Thus while in the zipper model the cooperativity of base pairing (i.e. the tendency for base pairs to open in contiguous segments, or in other words the existence of a nucleation penalty for the bubble) is put in "by hand", here it results from a local mechanism (the interplay between pairing and stacking). This idea is also present in the PB models, and indeed our approach is closest to the PB and NN models, but differing in the implementation of the underlying physics.

Stacking interactions in ss DNA were described years ago [16–18], but only 3 out of 16 stacking parameters have been measured precisely [2]. The major difficulty in the measurements is to separate the contributions of pairing and stacking interactions. As a result, the stacking interaction has not been incorporated into the DNA melting models as a separate degree of freedom. In practice, stacking was included in the NN model [3] as a correction to the pairing interaction parameter set. In the model below, we introduce stacking as a separate interaction with its own set of thermodynamic parameters (stacking enthalpy and entropy [3]). The thermodynamic parameters of the NN model can be calculated from these separate paring and stacking parameters, and appear then to be temperature dependent. We use the transfer matrix formalism to compute the partition function for our model, and compare to the experimental melting curves. In Sec. V-VII we extend this formalism to take into account the non local part of the loop entropies.

## II. STATES OF THE NN DIMER

Let us consider the NN model dimer $B_i B_{i+1} / B_i^* B_{i+1}^*$, $i = 1,\ldots,N-1$. The dimer has four DNA bases arranged into two anti-parallel strands $B_i B_{i+1}$ and $B_{i+1}^* B_i^*$ (listed from 5' to 3' end). The pairing interaction is between complementary bases $B_i$, $B_i^*$; the stacking interaction between adjacent bases $B_i$, $B_{i+1}$ and $B_{i+1}^*$, $B_i^*$ (considering stacking order from 5' to 3' end). The free energies of the $i$-th stackings between the bases $B_i$ and $B_{i+1}$ of the $B$-strand and bases $B_{i+1}^*$ and $B_i^*$ of the complimentary $B^*$-strand are $G_i^{St}$ and $G_i^{St*}$ respectively, while $G_i^P$ is the free energy of the $i$-th pairing between the bases $B_i$ and $B_i^*$. The free energy parameters result from the corresponding entropies and enthalpies, $G_i = E_i - T S_i$. To simplify the notation we will use the statistical weights: $U_i^P = \exp(-\boldsymbol{b} G_i^P)$, $U_i^{St} = \exp(-\boldsymbol{b} G_i^{St})$, and $U_i^{St*} = \exp(-\boldsymbol{b} G_i^{St*})$, where $\boldsymbol{b} = 1/T$. The model has a unique ground state - unmelted double helix with all bases paired and stacked - and many different exited states, even after complete strand dissociation. All thermodynamic parameters, i.e. pairing and stacking enthalpies, entropies, and free energies, are referred to the ground state. Therefore, unlike in most of the literature, we are dealing with positive energies and entropies of the excited states calculated with respect to the ground state. In a dimer base pairing is shared with the complementary bases (the stackings are not shared); therefore all pairings in the dimers will be counted with coefficient $\frac{1}{2}$. We suggest that if both pairing interactions in the dimer are "closed", then the stacking interactions are necessarily also closed, while if both pairings are "open", then the stackings may be closed or open, the opening of the stackings resulting in a free energy gain $G_i^{St}$ or $G_i^{St*}$. Thus in this description the free energy gain for complete unpairing of the NN dimer is:

$$G_i^{NN} = \left(G_i^P + G_{i+1}^P\right)/2 - T\left(\ln\left[1+\exp\left(-\boldsymbol{b}G_i^{St}\right)\right] + \ln\left[1+\exp\left(-\boldsymbol{b}G_i^{St*}\right)\right]\right). \qquad (1)$$

where the argument of the logarithm is the partition function expressing the fact that the unpaired dimer can be either stacked or unstacked. The enthalpy and entropy gain upon melting of the NN dimer can be easily calculated from Eq. (1), and it is evident that these effective thermodynamic parameters (which are the ones used in the NN model [3]) are now temperature dependent. We come back to this point later.

Each nearest neighbor dimer has 2 stacking degrees of freedom and 2 pairing degrees of freedom, which can each be open or closed, for a total of $2^4 = 16$ possible states. We introduce a local mechanism which gives rise to the cooperative behavior of the bubble, by implementing two local geometrical constraints. The first is that unstacking of two adjacent bases is only possible after unpairing of at least one of the bases. The second constraint requires at least one stacking of the dimer open if one pairing is open and one pairing closed. As a result, only 11 out of 16 states are admissible, Fig. 1. The geometrical origin of the constraints is that pairing interactions can prevent unstacking, and unpairing of exactly one out of two adjacent pairings requires spatial separation of the unpaired bases, which is impossible without al least one unstacking. Thus for instance Eq. (1) is applicable only for completely unpaired dimers, and not for partially melted dimers. These constraints require mandatory unstacking at the beginning of the melting fork, and therefore imply a penalty for opening a bubble.

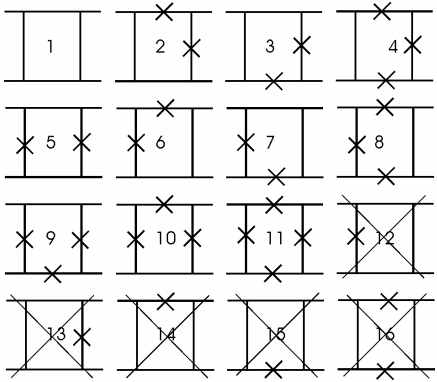

FIG. 1. Each NN dimer has two pairings (vertical lines) and two stackings (horizontal lines). The broken bonds are crossed. The horizontal lines represent the strands. There are sixteen states of the dimer; admissible states of the NN model dimer are the states from 1 to 11; the states from 12 to 16 are prohibited by the geometric constraints.

## III. TRANSFER MATRIX FORMALISM

It is convenient to write the partition function in the transfer matrix formalism; the mathematics is the same as for instance in the corresponding treatment of the 1-D Ising model of ferromagnetism [19] or the helix-coil transition in biopolymers [20,21]. Two adjacent NN dimers share one pairing interaction. Therefore, the transfer matrix technique for the model propagates the state of the pairing interactions only. The state of the $i$-th pairing is described by the two component covector (string) $X_i = (x_{i,0}, x_{i,1})$, $1 \leq i \leq N$. The first covector $X_1$ describes the boundary condition on the 5' end of the molecule. The covector component $x_{i,0}$ describes the partition function of the $i$-mer with the last pairing closed; $x_{i,1}$ describes the case of the last pairing open. The $2 \times 2$ transfer matrix $A_i$ transforms covector $X_i$ into covector $X_{i+1} = X_i A_i$ (there is no summation over $i$). The transfer matrix $A_i$ corresponds to the $i$-th NN dimer and contains the $i$-th stacking, the $i$-th stacking of the complementary strand, and the $i$-th and $i+1$ pairings.

To calculate the partition function of the entire molecule we need a boundary condition on the 3' end which can be characterized by the two dimensional vector (column) $Y_N$. In the following we will use free boundary conditions on the DNA molecule ends unless otherwise specified, i.e. all possible states of the first and the last base pairs are admissible. The vector $Y_i = (y_{i,0}, y_{i,1})^T$ corresponds to the $i$-th pairing (from the 5' end) and can be calculated recursively from the 3' end by: $Y_i = A_i Y_{i+1}$. The partition function of the whole DNA is the product of the 5' end boundary condition vector $X_1$, $(N-1)$ transfer matrixes corresponding to the NN dimers, and 3' end boundary condition vector $Y_N$:

$$Z_N = X_1 \left( \prod_{i=1}^{N-1} A_i \right) Y_N. \qquad (2)$$

The partition function can be also written as $Z_N = X_i Y_i$, there is no summation, and any $i$ from 1 to $N$ can be chosen. In the transfer matrix $A_i$:

$$A_i = \begin{pmatrix} a_{i,00} & a_{i,01} \\ a_{i,10} & a_{i,11} \end{pmatrix}. \qquad (3)$$

the value 0 of the index of the matrix elements corresponds to closed pairing, and 1 corresponds to open pairing, i.e. the matrix element $a_{i,00} = 1$ describes the ground state of the dimer with all pairings closed; $a_{i,01}$ describes the states with the first paring closed and the second pairing open; $a_{i,10}$ describes the states with the first pairing open and the second closed; and $a_{i,11}$ describes the states with both pairings open.

To write the explicit form of the matrix elements (3), note that $a_{i,00} = 1$ corresponds to diagram 1 in Fig. 1; $a_{i,01}$ corresponds to diagrams 2, 3, 4; $a_{i,10}$ corresponds to diagrams 6, 7, 8; and $a_{i,11}$ is the statistical weight corresponding to the free energy Eq. (1), and corresponds to diagrams 5, 10, 9, 11. The result for the transfer matrix is:

$$A_i = \begin{pmatrix} 1 & \sqrt{U_{i+1}^P} \left( U_i^{St} + U_i^{St*} + U_i^{St} U_i^{St*} \right) \\ \sqrt{U_i^P} \left( U_i^{St} + U_i^{St*} + U_i^{St} U_i^{St*} \right) & \sqrt{U_i^P U_{i+1}^P} \left( 1 + U_i^{St} + U_i^{St*} + U_i^{St} U_i^{St*} \right) \end{pmatrix} \qquad (4)$$

The contributions of the pairing free energies are counted with coefficients ½, or the square root of the statistical weights. The sums correspond to summing the different diagrams of Fig. 1 in the partition function; the products correspond to adding the free energies of pairing and stacking.

The first and the last dimers do not have adjacent dimers on their 5' or 3' ends respectively. The (co)vectors $X_1$ and $Y_N$ specifying the boundary conditions are multiplied by the first and the last transfer matrix in the partition function (2). Covector $X_1$ describes the first pairing interaction shared with the first dimer already described by the transfer matrix $A_1$, and vector $Y_N$ describes the last pairing shared with the last dimer already described by the transfer matrix $A_{N-1}$. We include ½ of the pairing interaction or the square root of the corresponding statistical weight into the 5' boundary covector $X_1$ and the 3' vector $Y_N$:

$$X_1 = \left(1, \sqrt{U_1^P}\right), \quad Y_N = \left(1, \sqrt{U_N^P}\right)^T. \tag{5}$$

To include in the model the entropy gain $S_D$ due to complete strand dissociation, we calculate the correction to the state with all pairings open:

$$Z_{N,D} = \left[\exp(S_D) - 1\right]\sqrt{U_1^P U_N^P} \prod_{i=1}^{N-1}\left[\sqrt{U_i^P U_{i+1}^P}\left(1 + U_i^{St}\right)\left(1 + U_i^{St*}\right)\right]. \tag{6}$$

The final partition function is the sum of the partition function (2) and the correction (6):

$$Z_{total} = Z_N + Z_{N,D}. \tag{7}$$

To calculate any local observables, for example the probabilities of unstacking or unpairing of some specific bond, we first calculate all $X$ and $Y$ vectors in advance. In the form of the partition function:

$$Z_N = X_i Y_i = x_{i,0} y_{i,0} + x_{i,1} y_{i,1} \tag{8}$$

(valid for any $i$), the first term is the partition function for the system with the $i$-th bp closed, the second term is the partition function for the $i$-th bp open. Thus the probability that the $i$-th bp is unpaired is:

$$P_{unpair}(i) = x_{i,1} y_{i,1} / Z_N . \tag{9}$$

Similarly, to calculate the probability that the $i$-th stacking of the primary strand is unstacked, we take the product $X_i A_i^1 Y_i$, where the matrix $A_i^1$ is derived from the matrix $A_i$ in (5) by keeping only the Fig. 1 diagrams corresponding to this unstacking, in this case diagrams 2, 4, 6, 8, 10, 11. Thus:

$$A_i^1 = \begin{pmatrix} 0 & \sqrt{U_{i+1}^P} \left( U_i^{St} + U_i^{St} U_i^{St*} \right) \\ \sqrt{U_i^P} \left( U_i^{St} + U_i^{St} U_i^{St*} \right) & \sqrt{U_i^P U_{i+1}^P} \left( U_i^{St} + U_i^{St} U_i^{St*} \right) \end{pmatrix} \tag{10}$$

and the required probability is:

$$P_{unstack}(i) = X_i A_i^1 Y_{i+1} / Z_N . \tag{11}$$

The average number of open bp is the sum $\sum_{i=1}^{N} P_{unpair}(i)$, and similarly for unstacking.

In summary, the difference between the $2 \times 2$ model (described by Eqs. (2) and (3)) and the existing standard form of the NN model [3] is that the partially melted dimers are considered specially. This internal structure of the dimers gives rise to a temperature dependence of the effective NN model parameters, expressed in Eq. (1). We show below that the $2 \times 2$ model is an improvement with respect to the NN model in describing oligomer melting curves. There is however still at least one important piece of physics missing from the model, namely a better estimate of the entropy of the bubbles (loops). In the present model, as in the NN model, the number of states increases exponentially with the length of the bubble, i.e. the entropy is linear in the bubble length. But in the real system the bubble is a loop, which results in a logarithmic correction to the entropy. The exact form of this correction (including excluded volume effects) has been addressed in the PS type models [5, 6]. It is possible to include these effects in a generalization of the present formalism which we present in Sec. V.

## IV. RESULTS FOR THE 2x2 MODEL

### A. Melting curves of oligomers

We compare the model to two oligomer sequences. The first oligomer sequence L13_2 used in the measurement is CG rich and forms no hairpins (CGA CGG CGG CGC G). The second sequence L60 is partially selfcomplementary and has an AT rich tract in the middle (CCG CCA GCG GCG TTA TTA CAT TTA ATT CTT AAG TAT TAT AAG TAA TAT GGC CGC TGC GCC). L60 can form a hairpin, but the contribution of hairpin states above the dissociation temperature is negligible [15].

For the experiments, synthetic oligomers were annealed in phosphate buffer saline (PBS) at an ionic strength of 50 mM, pH = 7.4 and the UV absorption curve $f$ was measured at 260 nm, in a 1 cm optical path cuvette (see Fig. 2 (A1) and (B1)).

To compare the model with the experimental data, the UV absorption is assumed to be a linear function of pairing and stacking:

$$f = (N\mathbf{a} \times (unpairings) + 2(N-1)\mathbf{d} \times (unstackings) + \mathbf{g})/C \qquad (12)$$

where $C$ is the concentration of the ds oligomer, measured in mmol, $\mathbf{a}$ and $\mathbf{d}$ are the molar extinction coefficients for unpairing and unstacking, measured in mmol$^{-1}$cm$^{-1}$ (the optical path being 1 cm). The UV curves were normalized so that $f = 1$ corresponds to strand dissociation. To measure the fraction of dissociated molecules $p$ for L60 we use a quenching technique which we have described before [22,23]. Briefly, to determine the dissociation curve $p$ (see Fig. 2 (A2)), the sample is heated at temperature $T_i$, then quenched to $\sim 0^\circ$C. Because the sequence L60 is partially self-complementary, molecules which were dissociated at temperature $T_i$ form hairpins after the quench. The relative number of hairpins (representing the fraction of dissociated molecules $p$ at temperature $T_i$) is determined by gel electrophoresis.

In Fig. 2 we display the melting curves from the experiments (symbols) and the model (continuous lines). In Fig. 2 (A2) and (B2) we show the fraction of dissociated molecules calculated from the model using the same parameter values as for the fit of the UV absorption curves. The sequence L13-2 is not self-complementary, so in Fig. 2 (B2) we show only the predicted dissociation curve.

For simplicity, in the fits the model was used with just one value for all stacking thermodynamic parameters, namely, stacking enthalpy and entropy of 4500 K and 12 respectively. Also, for L13-2 (which is nearly homogeneous) we used just one set of pairing enthalpy and entropy. Here and in the following we measure energies in degrees Kelvin and entropies in units of $k_B$. The high value of the dissociation entropy for L60 reflects a deficiency of the model: if we introduce a logarithmic bubble entropy in the form of Ref. [8], L60 can be fitted with the same dissociation entropy as for L13-2, while the other parameters of the fit change only about 5%. We cure this anomaly in Sect. V.

TABLE I. Values of the thermodynamic parameters which produce good fits to the experimental melting curves for L60 and L13-2, using the $2\times 2$ model. Energies are measured in Kelvins and entropies in units of $k_B$.

| Oligomer | $S_D$ | $E_{CG}^P$ | $S_{CG}^P$ | $E_{AT}^P$ | $S_{AT}^P$ |
|---|---|---|---|---|---|
| L60 | 34 | 5040 | 14 | 2800 | 7.4 |
| L13-2 | 9 | 5120 | 13.92 | N/A | N/A |

**B. Temperature dependence of the NN model parameters**

Fig. 3 shows the specific heat calculated from the partition function (2) for the oligomer L13-2. There is a gap in the specific heat on the two sides of the transition, which in this model is due to the residual stacking bonds breaking in the single stranded DNA (after strand separation). According to Fig. 3 the thermal capacity gain per dimer is about 78 at a temperature of 90°C, which is roughly consistent with the literature values reviewed in [24].

The thermodynamic parameters of the NN model can be calculated from the free energy of the completely unpaired dimer, Eq. (1). The enthalpy calculated from Eq. (1) is

$$E_i^{NN}(T) = \left(E_i^P + E_{i+1}^P\right)/2 + P_i^{St}(T)E_i^{St} + P_i^{St*}(T)E_i^{St*}, \qquad (13)$$

and the entropy is:

$$S_i^{NN}(T) = \left(S_i^P + S_{i+1}^P\right)/2 + \boldsymbol{b}\left(P_i^{St}(T)E_i^{St} + P_i^{St*}(T)E_i^{St*}\right)$$
$$+ \ln\left(\left[1+\exp\left(-\boldsymbol{b}G_i^{St}\right)\right]\left[1+\exp\left(-\boldsymbol{b}G_i^{St*}\right)\right]\right). \tag{14}$$

We plot these quantities in Fig. 4 for the oligomer L13-2 using the thermodynamic parameters from the UV spectrum fit of Fig. 2 (B1) and (B2).

## V. REFORMULATION OF THE MODEL TO INCLUDE THE POLAND – SHERAGA BUBBLE ENTROPIES

In the PS model the states of the DNA are classified in terms of non-interacting bubbles. The partition function counts all possible bubble states of the polymer and takes into account cooperative effects within the bubbles. Namely, the bubbles are characterized by the energy penalty of the broken bonds and the favorable entropy proportional to the bubble size, as well as an energy penalty for the creation of the melting forks. If the bubble is bounded by two ds segments, there is an extra entropy cost for terminating the random walk of the DNA single strands at the end of the bubble (i.e. making a closed loop). The energy penalty for the melting forks can, in our formalism, be absorbed into the statistical weights describing opening ($a_{01}$) and closure ($a_{10}$) of the bubble, in a sequence specific manner. The bubble entropy proportional to bubble size reflects the entropy gain upon breaking of the pairing interaction; we absorb it into the statistical weights describing open dimers ($a_{11}$). Then we are left with a bubble entropy of the form: $S^B(n) = -C_B \ln n$. The constant $C_B$ depends on the dimension of the random walk; its estimates can be found in an extensive literature [5, 6, 15, 17]. According to [3, 15, 17] the order of the phase transition at the strand separation temperature in the thermodynamic limit depends on the value of the constant $C_B$. If $C_B \leq 1$, the average size of the bubble is finite at any temperature and there is no phase transition. If $1 < C_B \leq 2$, the average size of size the bubble goes to infinity at the phase transition temperature and the phase transition is continuous. If $C_B > 2$, the average bubble size right before the phase transition is finite and the phase transition is of the first order.

The usual approach to the PS model is to consider the "bubble" as a particle and to calculate the grand partition function in the thermodynamic limit in terms of bubbles. The use of the grand partition function requires a variable number of DNA bases and the fugacity, which is not very transparent and non-constructive, i.e. the grand partition function cannot be used for the calculation of the partition function of an inhomogeneous DNA molecule of finite size, and is inefficient for calculating local quantities such as the probability of opening a bubble of a given size in a given place.

The local states of the NN model dimer described by the transfer matrix (3) and corresponding (co)vectors do not contain any information about the distance between bases once they have separated, or the bubble size. The long range bubble entropy penalty cannot be described by the $2\times 2$ matrices. Therefore, we extend the algorithm from the two-dimensional covectors $X_i$ (and corresponding vectors $Y_i$) of the $2\times 2$ model to multidimensional covectors. While the covector component $x_{i,0}$ still describes the partition function of the $i$-mer with the last pairing closed, the covector component $x_{i,n}$ ($n \geq 1$) is the partition function of the $i$-th mer with exactly $n$ open base pairs at the 3' end of the subsequence. For example, $x_{i,1}$ corresponds to the $i$th base pair open, with the $i-1$ base pair closed. For the $N$-mer the maximal size of the bubble is $N$ (the strand dissociated state) and the maximal size of covector $X_i$ is $N+1$.

We use the matrix elements of the transfer matrix $A_i$ introduced before to describe the local states of the NN dimer and parametrize the extended $(N+1)\times(N+1)$ transfer matrix $D_i$. The dimer with all base pairings closed is described by the matrix element $d_{i,00} = a_{i,00} = 1$. The dimer with the first base pair open and the second closed is characterized by the matrix element $d_{i,01} = a_{i,01}$. The fully open dimers are described by the matrix elements $d_{i,n\,n+1} = a_{i,11}$, where the index $n$ indicates how many bases are open in the 5' direction from the dimer. All bubble penalty effects (as well as the hairpin formation effects) are taken into account at the site of the bubble closure and incorporated into the corresponding matrix element $d_{i,n0} = U_i^B(n) a_{i,10}$, where $U_i^B(n) = \exp(S_i^B(n))$ (for $i > n$) are the statistical weights corresponding to the entropy penalty for closing the bubble, and $U_i^B(i) = 1$ since $n = i$ corresponds to the melting fork at the 5' end. All other elements of the transfer matrix $D_i$ are equal to zero. Explicitly, for the system of size N the covector $X_i$ has the form: $X_i = (x_{i,0}, x_{i,1}, \ldots, x_{i,N})$, where $x_{i,0}$ is the partition sum for the system of size $i$ with the last bp closed, while $x_{i,n}$ ($1 \leq n \leq i$) is the partition sum for the system of size $i$ with the last $n$ bp open. $x_{i,n} = 0$ for $n > i$, because the bubble cannot be larger than the size of the sequence. The recursion relations in this notation are:

$$x_{i+1,0} = a_{00} \, x_{i,0} + \sum_{n \neq 0} a_{i,10} \, U_i^B(n) \, x_{i,n}, \tag{15}$$

$$x_{i+1,n} = a_{i,11} \, x_{i,n-1}, \quad n \geq 1. \tag{16}$$

The transfer matrix of the model written in table form is thus:

$$D = \begin{pmatrix} 1 & a_{01} & 0 & 0 & \cdots \\ U^B(1)a_{10} & 0 & a_{11} & 0 & \cdots \\ U^B(2)a_{10} & 0 & 0 & a_{11} & \cdots \\ U^B(3)a_{10} & 0 & 0 & 0 & \cdots \\ \vdots & \vdots & \vdots & \vdots & \ddots \end{pmatrix}, \tag{17}$$

where the index $i$ is skipped for simplicity. The partition function of the oligomer can be calculated using Eq. (2) with the transfer matrix (17) and the free boundary conditions specified by the 5' covector and 3' vector:

$$X_1 = \left(1, \sqrt{U_1^P}, 0, \ldots, 0\right), \quad Y_N = \left(1, \sqrt{U_N^P}, \ldots, \sqrt{U_N^P}, \exp(S_D)\sqrt{U_N^P}\right)^T. \tag{18}$$

The square roots of the statistical weights in Eq. (18) count ½ of the pairing free energy not included into the first and the last NN dimers, which corresponds to the NN model [20] boundary corrections discussed in Sec. III. The last vector element $y_{N,N}$ corresponds to the bubble of size equal to the oligomer length, i.e. the state with all pairings open and separated strands. Therefore, the entropy gain upon strand dissociation $S_D$ is included in the $y_{N,N}$ vector component providing the partition function of the dissociated state. If we are not taking into consideration $S_D$ and the boundary condition effects, we can chose periodic boundary conditions. In the following we call the model described by Eqs. (2), (17), and (18) the $N \times N$ model.

To calculate the local observables, we proceed as in Sec. III (see Eq. (9) and following). For example, the probabilities that the $i$-th base pair is paired or unpaired are

$$P_{pair}(i) = x_{i,0} y_{i,0} / Z_N \quad \text{and} \quad P_{unpair}(i) = \sum_{n \neq 0} x_{i,n} y_{i,n} / Z_N . \tag{19}$$

We analyzed the experimental data of Fig. 2 with the $N \times N$ model, choosing the value of the bubble exponent $C_B = 2.1$ from [8]. The melting curves derived from the $N \times N$ model are essentially indistinguishable from the curves of the $2 \times 2$ model, upon small adjustments of the model's parameters. In Table I we list values of the thermodynamic parameters which produce melting curves for the 60-mer identical to the ones in Fig. 2, using the $2 \times 2$ and $N \times N$ models. The parameters were adjusted "by hand", because there are too many parameters to perform a global fit. The stacking parameters were kept the same for both models. The introduction of the bubble exponent makes little difference for the 13-mer, whereas for the 60-mer the dissociation entropy $S_D$ in the $N \times N$ model is 9 instead of 34, i.e. the same as for the 13-mer (see Table I). Thus the $N \times N$ model cures an anomaly of the $2 \times 2$ model, since we expect the dissociation entropy to depend on DNA concentration, not DNA length.

TABLE II. Values of the thermodynamic parameters which produce good fits to the experimental melting curves for L60, using the $2 \times 2$ and $N \times N$ models. Energies are measured in Kelvins and entropies in units of $k_B$.

| Model | $S_D$ | $E_{CG}^P$ | $S_{CG}^P$ | $E_{AT}^P$ | $S_{AT}^P$ |
|---|---|---|---|---|---|
| $2 \times 2$ model | 34 | 5040 | 14 | 2800 | 7.4 |
| $N \times N$ model | 9 | 5000 | 14.2 | 2700 | 7.6 |

The advantage of our model is the correspondence of the algorithm implementation to the physical structure of the DNA molecule. Each transfer matrix of the model corresponds to the NN dimer and contains the NN dimer data in a clear and explicit form. The boundary condition vectors in the algorithm correspond to the boundary corrections of the NN model and contain the NN model initiation parameters. The long range bubble effects are incorporated in our transfer matrix and counted in the partition function calculation at the site of the bubble closure (or opening for the reverse sequence and the transposed transfer matrixes). Thus this long range interaction can be efficiently considered within the transfer matrix technique as a local effect on the new extended states of the model.

All local observables and local correlations can be easily derived from the partition function of the model in the arbitrary sequence case. The other advantage of the transfer matrix technique is an easy way of getting the thermodynamic limit for homogeneous sequences, without using the grand partition function. In the next sections we study the thermodynamic limit of the model in the homogeneous case.

## VI. EIGENVALUES AND EIGENVECTORS FOR THE $N \times N$ MODEL

In the following we assume that the bubble entropy penalty is proportional to the logarithm of the bubble size: $S^B(n) = -C_B \ln n$, and the Boltzmann exponent of the bubble entropy penalty is

$$U^B(n) = \exp\left(S^B(n)\right) = n^{-C_B}. \tag{20}$$

Decomposing the determinant of the eigenvalue equation, $\det(D - I\mathbf{1}) = 0$, into the minors of the first column we get

$$(1-I)M_{00} + a_{10}\sum_{n=1}^{N}(-1)^n U^B(n)M_{n0} = 0, \tag{21}$$

where $M_{00} = (-I)^N$, and $M_{n0} = a_{01}a_{11}^{n-1}(-I)^{N-n}$, and the eigenvalue equation is

$$I^N\left[(1-I) + \frac{a_{01}a_{10}}{a_{11}}\sum_{n=1}^{N}\left(\frac{a_{11}}{I}\right)^n n^{-C_B}\right] = 0. \tag{22}$$

If $C_B = 0$, there is no bubble entropy penalty, and the $N \times N$ model reduces to the $2 \times 2$ model. Eq. (22) is a polynomial equation of the power $N+1$ and it has $N+1$ solutions $l_i$, $i = 0,1,\ldots N$. Because all the coefficients in Eq. (22) are positive, except the coefficient of the term with the highest power of $l$, only one solution is positive and greater than 1. In the following we will use the notation $l$ only for the largest positive eigenvalue. In the thermodynamic limit $N \to +\infty$, the sum is a polylogarithm function (also known as Jonquière's function), $\sum_{n=1}^{+\infty} \left(\frac{a_{11}}{l}\right)^n n^{-C_B} = \text{Li}_{C_B}\left(\frac{a_{11}}{l}\right)$, and the equation takes the form

$$l = 1 + \frac{a_{01}a_{10}}{a_{11}} \text{Li}_{C_B}\left(\frac{a_{11}}{l}\right). \tag{23}$$

The polylogarithm Eq. (23) has only one positive solution, since the left side of the equation is an increasing and the right side a decreasing function of $l$. Before the phase transition $l > a_{11}$, and $a_{11} = l$ at the phase transition (see below).

The transfer matrix is non-Hermitian, and the left and the right eigenvectors are different, but the eigenvalue equation and the eigenvalues are still the same. The left eigencovector $V \equiv (v_0, v_1, v_2, \ldots)$ equations for the transfer matrix (17) are

$$v_0 + \sum_{n=1}^{+\infty} v_n U^B(n) a_{10} = l v_0, \tag{24}$$

$$v_0 a_{01} = l v_1, \tag{25}$$

$$v_n a_{11} = l v_{n+1}, \quad n \geq 1. \tag{26}$$

From Eqs. (25) and (26) one can choose:

$$v_0 = \frac{a_{11}}{a_{01}}, \qquad v_n = \left(\frac{a_{11}}{\bm{l}}\right)^n, \quad n \geq 1, \tag{27}$$

and the Eq. (24) becomes equivalent to Eq. (23).

The right eigenvector $W$ equations are

$$w_0 + a_{01} w_1 = \bm{l}\, w_0, \tag{28}$$

$$U^B(n) a_{10} w_0 + a_{11} w_{n+1} = \bm{l}\, w_n, \quad n \geq 1. \tag{29}$$

The eigenvalue Eq. (23) can be derived by exclusion $w_n$, $n \geq 1$ from the Eq. (28) using Eq. (29). The eigenvector elements can be calculated recursively from the Eqs. (28) and (29). Let us define the scalar product of a covector $V$ and a vector $W$ as $(V,W) = \sum_{n=0}^{N} v_n w_n$. Unlike the Hermitian case, the basis sets in vector and covector spaces are different, and the scalar product can be calculated between covector and vector only.

## VII. PARTITION FUNCTION IN THE THERMODYNAMIC LIMIT

We calculate the partition function of the homogeneous $N$-mer with the boundary conditions $X_1 = \Phi$, $Y_N = \Phi^T$, where $\Phi \equiv (1,0,0,\ldots)$, i.e. the first and the last base pairs are paired. The boundary conditions can be decomposed in the basis of eigencovectors and eigenvectors as

$$X_1 = \sum_i \frac{(X_1, W_i) V_i}{(V_i, W_i)}, \text{ and } Y_N = \sum_i \frac{W_i (V_i, Y_N)}{(V_i, W_i)}. \tag{30}$$

In the thermodynamic limit, only the largest eigenvalue $\bm{l}$ with eigencovector $V$ and eigenvector $W$ are needed for the partition function calculation. This statement is not true in the case of an infinite number of eigenvalues or in the presence of a continuum spectrum. Instead, one can verify that any local perturbation of the local boundary condition is growing (in the transfer matrix multiplication) as the eigen(co)vector with the highest eigenvalue or slower. The highest eigenvalue eigenvector component of the boundary conditions can be found from the scalar product. The partition function of the DNA $N$-mer in the large $N$ limit is:

$$Z_N = \Phi D^{N-1} \Phi^T = \frac{(\Phi,W)V}{(V,W)} D^{N-1} \Phi^T = \frac{\mathbf{1}^{N-1}(\Phi,W)(V,\Phi^T)}{(V,W)}, \quad (31)$$

where $(\Phi,W) = w_0$, $(V,\Phi^T) = v_0$, and the scalar product $(V,W)$ can be calculated from the recursive Eqs. (27) – (29) as

$$(V,W) = w_0 v_0 \left[ 1 + \frac{a_{01} a_{10}}{\mathbf{1} a_{11}} \text{Li}_{C_B - 1}\left(\frac{a_{11}}{\mathbf{1}}\right) \right]. \quad (32)$$

We calculate the probability of some configuration of the DNA molecule by implementing extra local constraints on the states counted in the partition function. The probability of the configuration will be equal to the ratio of the partition function with the extra constraints divided by the unrestricted partition function (31). The procedure for implementing any local constraint is straightforward in the transfer matrix formalism.

The partition function of the $N$-mer with the $M$-th base pair closed (located far away from the DNA ends) is:

$$Z_{N,\text{pair }M} = \left[ \Phi D^{M-1} \Phi^T \right]\left[ \Phi D^{N-M} \Phi^T \right]$$
$$= \left[ \frac{(\Phi,W)V}{(V,W)} D^{M-1} \Phi^T \right]\left[ \frac{(\Phi,W)V}{(V,W)} D^{N-M} \Phi^T \right] = \frac{\mathbf{1}^{N-1}(\Phi,W)^2 (V,\Phi^T)^2}{(V,W)^2}. \quad (33)$$

The fraction of unpaired bases in the whole DNA is equal to the probability of the $M$-th base pair been unpaired:

$$P_{\text{pair}} = \frac{Z_{N,\text{pair }M}}{Z_N} = \frac{w_0 v_0}{(V,W)} = \left[ 1 + \frac{a_{01} a_{10}}{\mathbf{1} a_{11}} \text{Li}_{C_B - 1}\left(\frac{a_{11}}{\mathbf{1}}\right) \right]^{-1}. \quad (34)$$

The ratio of the partition functions with open $M$-th NN dimer and the unconstrained partition function (32) is equal to the probability of a bubble of arbitrary size opening at the dimer position $M$:

$$P_{opening} = \frac{a_{01} w_1 v_0}{\boldsymbol{l}(V,W)} = \left[1 - \boldsymbol{l}^{-1}\right]\left[1 + \frac{a_{01} a_{10}}{\boldsymbol{l} a_{11}} \text{Li}_{C_B-1}\left(\frac{a_{11}}{\boldsymbol{l}}\right)\right]^{-1}, \quad (35)$$

where, according to Eq. (28), $w_1 = (\boldsymbol{l} - 1) w_0 / a_{01}$. The density of bubbles of size $n$ is equal to the probability that any particular location (far away form the boundaries) is the site of the closure or opening of a bubble of size $n$:

$$P_{bubble}(n) = \frac{U^B(n) a_{10} v_n w_0}{\boldsymbol{l}(V,W)} = n^{-C_B}\left(\frac{a_{11}}{\boldsymbol{l}}\right)^n \left[\frac{\boldsymbol{l} a_{11}}{a_{01} a_{10}} + \text{Li}_{C_B-1}\left(\frac{a_{11}}{\boldsymbol{l}}\right)\right]^{-1}, \quad (36)$$

and the average size of the bubble can be calculated from the bubble size distribution (36):

$$<n> = \frac{\sum_{n=1}^{+\infty} n P_{bubble}(n)}{\sum_{n=1}^{+\infty} P_{bubble}(n)} = \frac{\text{Li}_{C_B-1}\left(\frac{a_{11}}{\boldsymbol{l}}\right)}{\text{Li}_{C_B}\left(\frac{a_{11}}{\boldsymbol{l}}\right)}. \quad (37)$$

The average size of the bubble still depends on the matrix elements $a_{01}$ and $a_{10}$, because the eigenvalue $\boldsymbol{l}$ depends on them.

### A. The phase transition

The phase transition occurs if the discrete spectra eigenvector gets divergent norm, i.e. $a_{11} = \boldsymbol{l}$, and the contribution of the closed states $v_0^2$ to the partition function vanishes in comparison with the infinite norm of all separated states $\sum_{i=1}^{+\infty} v_i^2$. If the matrix elements of $A$ are provided explicitly, we can solve Eq. (23) with $a_{11} = \boldsymbol{l}$ to find the phase transition temperature:

$$a_{11} = 1 + \frac{a_{01}a_{10}}{a_{11}} z(C_B). \tag{38}$$

We consider the Riemann Zeta function $z(C_B) = \text{Li}_{C_B}(1)$ divergent, if its argument is less than 1. In that case Eq. (38) has no solution, $a_{11} < 1$ for any temperature, and there is no phase transition.

The probability of the bubble of size $n$, Eq. (36), is proportional to the product of $v_n = (a_{11}/1)^n$ and the corresponding matrix elements $U^B(j)a_{10} \propto n^{-C_B}$, i.e. it is the product of an exponential and power functions of $n$. At the phase transition temperature $a_{11} = 1$ and the probability of the bubble size $n$ becomes a power law $\propto n^{-C_B}$ and the average bubble size is the ratio of the Riemann Zeta functions:

$$<n> = \frac{z(C_B - 1)}{z(C_B)}. \tag{39}$$

If $1 < C_B < 2$, the average size of the bubble is going to infinity near the temperature of the phase transition and the phase transition is continuous. If $C_B \geq 2$, the average bubble size at the phase transition is finite, and the phase transition is of the first order. The order of the phase transition was calculated in the literature [4, 25] using the grand partition function. However the transfer matrix technique allows us to calculate the bubble distribution explicitly.

### B. Critical exponent for the continuous phase transition, $1 < C_B \leq 2$

We use the following properties of the Polylogarithm function: $\frac{d}{dx}\text{Li}_{C_B}(x) = 1 + \text{Li}_{C_B - 1}(x)$, and $\text{Li}_C(x) \propto (1-x)^{C-1}$ for $x \to 1-0$ and $C \leq 1$. Thus Eq. (34) becomes

$$P_{pair} \propto (1 - a_{11})^{2 - C_B} \tag{40}$$

near the melting temperature $T_M$. Using Eq. (23) we decompose $\mathbf{1}$ into a Taylor series at the phase transition point. It is easy to check that at the phase transition $\frac{d\mathbf{1}}{da_{11}} = 1$, and

$$[T_M - T] \propto [a_{11}(T_M) - a_{11}(T)] \approx [\mathbf{1}(T_M) - \mathbf{1}(T)] \propto [\mathbf{1}(T) - a_{11}(T)]^{C_B - 1}. \quad (41)$$

Combining Eqs. (40) and (41) we obtain a critical exponent for the phase transition, as in [26]:

$$P_{pair} \propto [T_M - T]^{(2 - C_B)/(C_B - 1)}. \quad (42)$$

## VIII. CONCLUSIONS

We propose a model for the temperature driven melting of DNA secondary structure where pairing and stacking interactions between bases are explicitly introduced as distinct degrees of freedom. The geometrical constraints imposed by the structure of the double helix (the ground state), such as for example the fact that one isolated unpairing also forces at least one unstacking, are implemented through restrictions in the possible states of the NN dimer. This local mechanism results in the appearance of longer range correlations responsible for the cooperative opening of bubbles bounded by ds segments, as observed in experiments [23]. The model describes well the melting curves of oligomers in the entire temperature range, including after strand separation, where the effect of residual stacking is not captured by previous models. In addition, the model explains the observed temperature dependence of the NN model parameters, and the gap in the specific heat. The model in its simplest form treats the ss segments as ideal random walks (number of states increasing exponentially with length), similarly to the NN model. An extension of the same transfer matrix formalism properly accounts for the entropy of the closed loops. The advantage of our algorithm is the simple structure of the transfer matrix. The structure of the transfer matrixes directly reflects the real structure of the NN model dimers and the stacking interaction of the double and single stranded DNA can be taken into account. The matrix elements are the statistical weights of the dimer states, and the boundary condition vectors correspond to the NN model boundary corrections. The bubbles are counted in the transfer matrix at the site of their closure, and become local objects. The size of the transfer matrix is the length of the polymer, but most of the transfer matrix elements are zero. The algorithm complexity grows quadratically and computer memory grows linearly with DNA length, like for the classic Poland's algorithm [5]. Finally, the transfer matrix technique allows to easily calculate the homogeneous thermodynamic limit of the model, without use of the grand partition function.

## ACKNOWLEDGEMENTS

We thank Prof. Joseph Rudnick (UCLA) and Mikhail G. Ivanov (MIPT, Moscow, Russia) for discussions of the model, and Michael Entin (Microsoft) for advice on the parameter search.

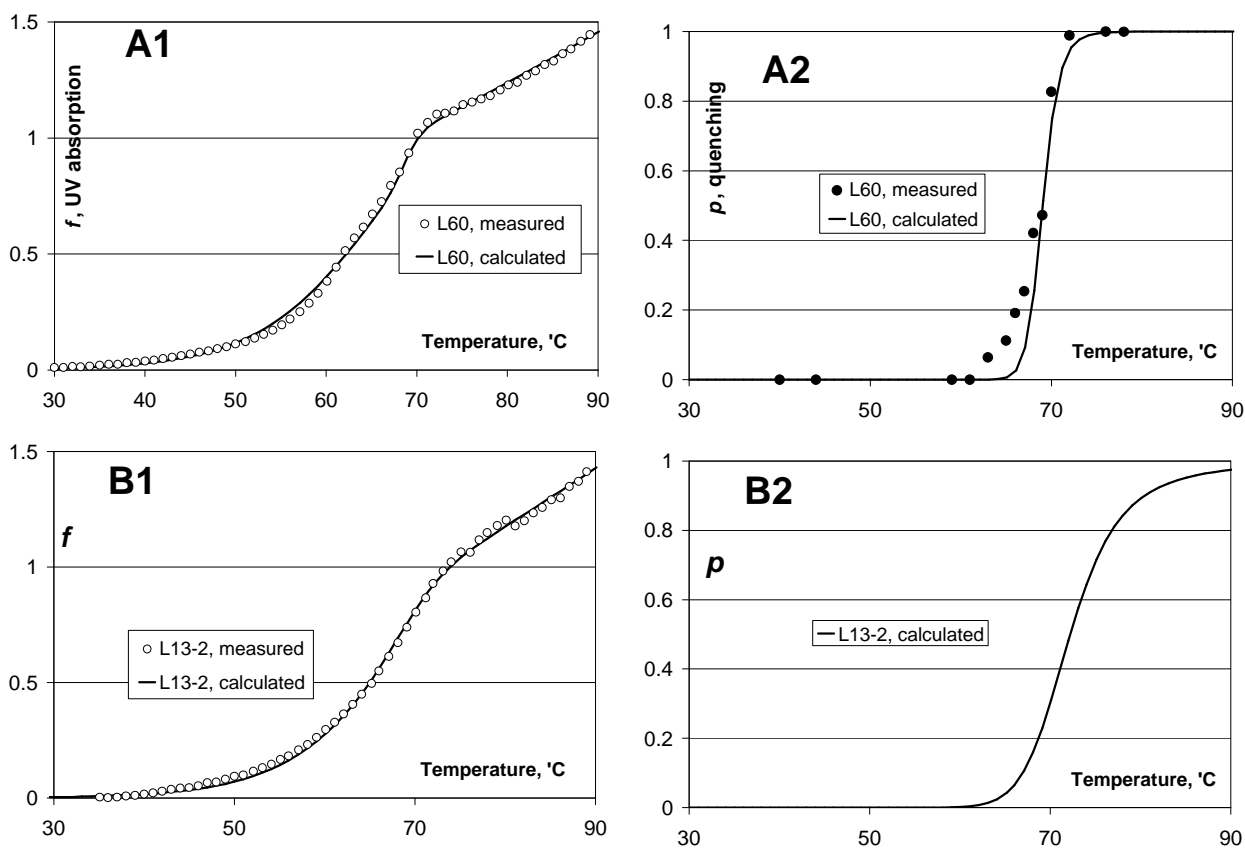

FIG. 2. (A1, B1): normalized UV absorption spectra *f* measured at 260 nm for the L60 and L13-2 ds DNA oligomers. The experiments are the circles; the $2\times 2$ model is the solid line.

(A2, B2): measured and predicted dissociation curves *p* for L60, and predicted dissociation curve for L13-2. The measurements in (A2) (filled circles) were obtained from the quenching method. The $2\times 2$ model is plotted using the same parameter values as in (A1), (B1). The source of the discrepancy between the experimental data and the model in A2 is not clear, but possibly it is related to the hairpin states of the 60mer, which are not counted in the partition function of the model.

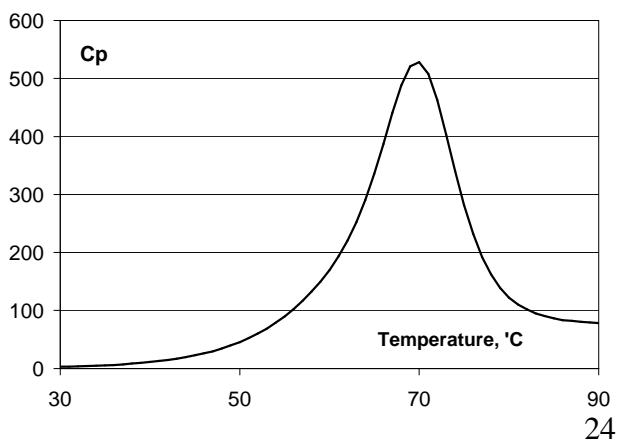

FIG. 3. Thermal capacity of the DNA (per NN dimmer, in units of $k_B$) calculated from the $2\times 2$ model partition function of the 13-mer. The residual thermal capacity after strand dissociation is due to the residual stacking.

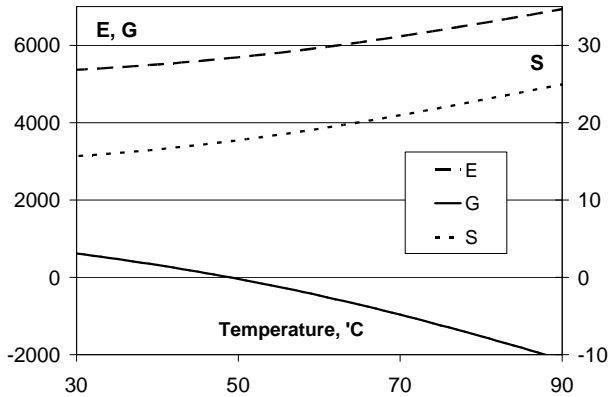

FIG. 4. Enthalpy *E* (long dash), entropy *S* (short dash) and free energy *G* (solid) gain upon NN dimer opening, calculated for L13-2 from Eqs. (1), (13), and (14). Energies are measured in Kelvins (left hand scale), and the entropy in units of $k_B$ (right hand scale).